\documentclass[
  aps,
  prl,
  reprint,
  amsmath,
  amssymb,
  footinbib,
  longbibliography,
  superscriptaddress
]{revtex4-2}

\usepackage{bm}
\usepackage{mathtools}
\usepackage[dvipsnames]{xcolor}
\usepackage{graphicx}

\usepackage[bookmarksnumbered=true,bookmarksopen=true]{hyperref}
 \hypersetup{colorlinks,%
             linkcolor=NavyBlue, %
             citecolor=PineGreen, %
             urlcolor=PineGreen}

\graphicspath{{fig/}}

\def\bge{\begin{equation}}
\def\ede{\end{equation}}
\def\bga{\begin{aligned}}
\def\eda{\end{aligned}}
\def\bgb{\begin{bmatrix}}
\def\edb{\end{bmatrix}}
\def\bgp{\begin{pmatrix}}
\def\edp{\end{pmatrix}}
\def\bgm{\begin{matrix}}
\def\edm{\end{matrix}}
\def\bgs{\begin{subequations}}
\def\eds{\end{subequations}}
\newcommand{\order}[1]{\mathcal{O}({#1})}
\def\di{{\mathrm{d}}}

\def\Di{{\mathcal{D}}}

\newcommand{\n}{\nonumber}
\def\T{\mathcal{T}}

\def\pd{\partial}

\def\la{\langle}\def\ra{\rangle}

\def\to{\rightarrow}

\def\ii{\mathrm{i}}

\def\be{\beta}

\def\de{\delta}

\def\si{\sigma}

\def\aa{\mathsf{a}}
\def\bb{\mathsf{b}}

\def\H{{\mathcal{H}}}

\def\J{{\mathcal{J}}}

\newcommand{\sA}{\mathsf{A}}
\newcommand{\nv}{\widetilde{\nu}} 
\newcommand{\M}{\widetilde{\mathcal D}} 

\newcommand{\Ima}{\operatorname{Im}}
\newcommand{\sech}{\operatorname{sech}}

\newcommand{\wt}[1]{\mkern 2mu \widetilde{\mkern -2mu #1 \mkern -2mu}\mkern 2mu}

\renewcommand{\FR}[2]{\displaystyle\frac{\,{#1}\,}{#2}}
\newcommand{\fr}[2]{\mbox{$\frac{\,{#1}\,}{#2}$}}

\begin{document}
\hbadness=10000

\title{Triple-Massive-Exchange Bispectrum of Quasi-Single-Field Inflation}

\author{Xingang Chen}
\email{xingang.chen@cfa.harvard.edu}
\affiliation{Institute for Theory and Computation, Harvard-Smithsonian Center for Astrophysics, 60 Garden Street, Cambridge, MA 02138, USA} 

\author{Zhong-Zhi Xianyu}
\email{zxianyu@tsinghua.edu.cn}
\affiliation{Department of Physics, Tsinghua University, Beijing 100084, China}

\begin{abstract}

The triple-exchange bispectrum of the scalar quasi-single-field inflation model has resisted analytic evaluation across its full parameter space since its advent, despite rapid progress in cosmological collider physics. In this Letter, we solve it for general mass and mixing strength as a convergent double series, by applying the Mellin-Barnes representation directly to the mixed propagator between the massless and massive modes. The result is exact, non-perturbative, and physically transparent; the method is broadly applicable.

\end{abstract}

\maketitle

\noindent\textbf{\textit{Introduction.---}} 
The program of cosmological collider (CC) physics, probing particle physics at the energy scale of inflation, is gaining increasing attention since its inception \cite{Chen:2009we,Chen:2009zp,Baumann:2011nk,Noumi:2012vr,Arkani-Hamed:2015bza}, attracting efforts from particle model building \cite{Chen:2016uwp,Chen:2016hrz,Lee:2016vti,Kumar:2017ecc,Hook:2019vcn,Bodas:2024hih,Sundrum:2026ghf}, data analysis \cite{Planck:2013wtn,Cabass:2024wob,Sohn:2024xzd,Goldstein:2024bky,Suman:2025vuf,Suman:2025tpv,Kumar:2026ogn,Kumar:2026dih,Philcox:2026tjj,Green:2026yev}, to amplitude technique development \cite{Baumann:2022jpr}. 

In terms of particle content, models in this program contain spectator particles with masses of order the Hubble parameter or less and are known as quasi-single-field inflation (QSFI) models. The first example of a QSFI model features a Hubble-scale scalar spectator \cite{Chen:2009we,Chen:2009zp,Chen:2012ge}. 
Besides serving as a prototypical model of CC physics and quantum primordial standard clocks \cite{Chen:2015lza}, this model also belongs to one of several classes of models that naturally generate large non-Gaussianities, as classified in \cite{Wang:2019gbi}. In particular, the triple-exchange bispectrum of this model has long been recognized as a channel for large leading non-Gaussianity and a promising target for observations \cite{Sefusatti:2012ye,Planck:2013wtn}.

One of the major goals of CC physics is to achieve analytic and numerical control over the correlators predicted by various models. This is challenging because correlators involving massive propagators in inflationary spacetime are significantly harder to compute than their flat-spacetime counterparts.

Several numerical strategies with varying degrees of efficiency exist for this calculation. The combination of mixed propagators and Wick-rotated time integrals is itself a highly efficient numerical method \cite{Chen:2009zp,Chen:2017ryl,Kumar:2026dih}. Alternatively, the calculation can be performed using differential-equation-based \cite{Werth:2023pfl,Pinol:2023oux,Werth:2024aui} or integration-by-parts-based techniques \cite{Chen:2023iix,Chen:2026iwf}. Numerical methods have also been applied to signal searches in observational data \cite{Sohn:2024xzd,Suman:2025vuf,Suman:2025tpv,Kumar:2026ogn,Kumar:2026dih,Philcox:2026tjj,Cabass:2024wob,Goldstein:2024bky,Green:2026yev}.

Meanwhile, analytic control of correlators, whether in closed form or through convergent series, remains indispensable for gaining physical insights and guiding data analysis templates. Substantial efforts have been devoted to the analytic study of cosmological correlators, aiming to obtain explicit analytic solutions and understand their analytic structure \cite{Arkani-Hamed:2018kmz,Baumann:2019oyu,Sleight:2019mgd,Sleight:2019hfp,Pimentel:2022fsc,Jazayeri:2022kjy,Qin:2022fbv,Qin:2023ejc,Xianyu:2023ytd,Aoki:2024uyi,Liu:2024str,Xianyu:2025lbk,Werth:2024mjg,Belrhali:2026rkn,Belrhali:2026jqe,Belrhali:2026act,Grafe:2026avi}.

Surprisingly, despite the remarkable progress and many fruitful results in CC physics, full analytic evaluation over the triple-exchange bispectrum of the first QSFI model over its general parameter space remains an open challenge. The difficulty mainly arises from the many internal massive modes and the relatively few independent kinematic variables. Recently, some interesting partial progress has been made, including a convergent-series solution in part of the parameter space \cite{Belrhali:2026uxn} and a closed-form solution but with its defining series non-converging for physical kinematics \cite{Xianyu:2025lbk}. Meanwhile, exact mode functions at arbitrary mixing strength have been found \cite{Huenupi:2026abj,Huenupi:2026aqc,Wang:2026lff,Pinol:2026xnl,Pinol:2026hey}, raising an even more challenging question of analytically solving the non-perturbatively mixed bispectrum of this model.

In this Letter, we resolve both challenges through a different approach that combines two existing techniques: the mixed propagator and the partial Mellin-Barnes (PMB) representation \cite{Chen:2017ryl,Sleight:2019mgd,Sleight:2019hfp,Qin:2022lva,Qin:2022fbv}. 
With the mixed propagator, the process involves only one time integral; with PMB, all special functions in the propagator is resolved into power factors, and the Mellin integrals can be easily carried out as converging series thanks to the meromorphic nature of the mixed propagator. Crucially, with a prescription on the summation order, the resulting series converges for all physical kinematics with nonzero momenta, except for the equilateral point,  which is covered by a simple variation of the original series. Below, we report the details.

We work with the dS background with metric $\di s^2=a^2(-\di\tau^2+\di\bm x^2)$ where the conformal time $\tau\in(-\infty,0)$, the scale factor $a=-1/(H\tau)$, and we set Hubble parameter $H=1$. 

\noindent\textbf{\textit{The triple-exchange bispectrum of a scalar QSFI model.---}} The model introduced in Refs.~\cite{Chen:2009we,Chen:2009zp} contains an inflaton $\phi$ and a spectator scalar field $\si$ of mass $m\sim \order{H}$. The inflaton $\phi=\phi_0+\varphi$ has a spatially uniform and slowly rolling background $\phi_0(\tau)$ and a nearly massless fluctuation $\varphi(\tau,\bm x)$.

This scalar QSFI model may be realized by Lorentz covariant models in various ways \cite{Chen:2009we,Chen:2009zp,Baumann:2011nk,Noumi:2012vr,Reece:2022soh,Xianyu:2025lbk}, but at the level of fluctuations, they generally lead to following interactions:
\begin{equation}
 S_{\rm int}=\int \di\tau \di^3\bm x\,
 \Big[a^3\rho\varphi'\sigma-\fr16 a^4\lambda_3\sigma^3\Big]+\cdots.
 \label{eq_int}
\end{equation}
where $\varphi'\equiv\pd_\tau\varphi$ mixes with $\si$ with a strength $\rho$. There are also other interactions denoted by $\cdots$ but they generally do not lead to a bispectrum as large as the triple-exchange one. The mixing $\rho$ can be either large or small. Below, we first focus on the weakly mixing case with small $\rho$, take the $\si$'s mass $m>3/2$, define $\wt\nu\equiv\sqrt{m^2-9/4}$, and assume Bunch-Davies states for both fields. Extension to nonperturbative mixing with large $\rho$ is straightforward and will be addressed shortly. The $0<m\leq 3/2$ case is the usual analytic continuation.

In typical QSFI models, the single-, double- and triple-exchange processes coexist, but the triple exchange is often the dominant one \cite{Chen:2009zp,Chen:2018sce}. So we focus on the triple exchange only. Following the convention of \cite{Chen:2010xka}, we define a dimensionless shape function for the curvature bispectrum, which is related to the 3-point inflaton correlator via:
$\mathcal S =- (k_1k_2k_3)^2/(2\pi P_\zeta^{1/2})\times
 \langle\varphi_{\bm k_1}\varphi_{\bm k_2}\varphi_{\bm k_3}\rangle'$,
where $\la\cdots\ra'$ means the momentum conservation factor $(2\pi)^3\de(\bm k_1+\bm k_2+\bm k_3)$ neglected. This conservation limits physically reachable kinematics to be all nonnegative $k_1,\cdots,k_3$ ($k_i\equiv|\bm k_i|$) with triangle inequality satisfied. The shape function depends on $k_i$ only through two independent momentum ratios which we choose to be $r_1\equiv k_1/k_3$ and $r_2\equiv k_2/k_3$. Without loss of generality, we choose to work in the physical region where $0<k_1\leq k_2\leq k_3$ so that $0<r_1\leq r_2\leq 1$ and $r_1+r_2\geq 1$. Also, $P_\zeta=1/(4\pi^2\dot\phi_0^2)\simeq 2\times 10^{-9}$ is the measured scalar power around the CMB scale where the curvature fluctuation $\zeta\simeq -\varphi/\dot\phi_0$ \cite{Planck:2018jri}. 

\begin{figure}
\includegraphics[width=0.2\textwidth]{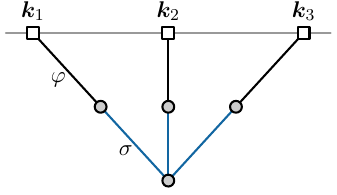}
\caption{The triple-massive-exchange diagram. The top horizontal line shows the end of inflation surface $\tau=0$ on which we compute the correlator $\la\varphi_{\bm k_1}\varphi_{\bm k_2}\varphi_{\bm k_3}\ra'$. The black (blue) lines denote bulk-to-boundary (bulk-to-bulk) propagators $G_\aa$ of $\varphi$ ($D_{\aa\bb}$ of $\si$). }
\label{fig_triple_exchange}
\end{figure}

Then, with interactions in (\ref{eq_int}), the leading contribution to the inflaton 3-point correlator comes from a process of triple massive exchange, as shown in Fig.\;\ref{fig_triple_exchange}. 

The diagrammatic rules for computing this process was given in \cite{Chen:2017ryl}, which lead to the following time integral for the shape function:
\begin{align}
 \mathcal S 
 =&-\FR{\rho^3\lambda_3}{\pi P_\zeta^{1/2}}\FR{\Ima\J(r_1,r_2)}{r_1r_2},
\end{align}
where the function $\J$, also called the bispectrum for convenience, is computed via the following integral:
\begin{align}
 \J(r_1,r_2)=&k_1^3k_2^3  \int_{-\infty}^0\frac{\di\tau}{\tau^4}
       \prod_{i=1}^3\mathcal{G}_+(k_i;\tau) . 
\end{align}
Here, following \cite{Chen:2017ryl}, we have introduced the \emph{mixed propagator}, which is simply the mixing of a $\varphi$-line and a $\sigma$-line in Fig.\;\ref{fig_triple_exchange}. It is computed by the following time integral of the $\varphi$-propagator $G_\bb$ and $\si$-propagator $D_{\aa\bb}$; see \cite{Chen:2017ryl} for details. 
\begin{align}\label{eq:mixed} 
 \mathcal G_\pm(k;\tau)&=\ii \sum_{\bb=\pm}\bb\int_{-\infty}^0\FR{\di\tau'}{-\tau'^3}
 D_{\pm\bb}(k;\tau,\tau')\partial_{\tau'}G_\bb(k;\tau')
\n\\ 
 &\equiv \frac{1}{k^3}\Di_\pm(-k\tau). 
\end{align}
In the last line, we also defined a dimensionless mixed propagator $\Di_\pm(z)$. Below we write $\mathcal{D}\equiv\mathcal{D}_+$ and then the bispectrum $\J$ can be expressed as
\begin{align}
 \J(r_1,r_2)&=\int_0^\infty\frac{dz}{z^4}
       \Di(r_1z)\Di(r_2z)\Di(z).
\label{eq_BispectrumTimeIntegral}
\end{align}
Therefore, the study of the bispectrum $\J$ boils down to a problem of finishing the integral (\ref{eq_BispectrumTimeIntegral}).

\noindent\textbf{\textit{Mixed propagator.---}}
The mixed propagator $\Di(z)$ in (\ref{eq:mixed}) can be directly integrated as was done in \cite{Chen:2017ryl}. Here we quote the result in an equivalent form:
\begin{align}
 \Di(z)={}&\sum_{\aa=\pm} \be_\aa(\wt\nu)\text{J}_{\ii\aa \widetilde{\nu}}(z)\,z^{3/2} 
 \nonumber\\[-2pt]
 &-\frac{z^2e^{-\ii z}}{2(1/4+\widetilde{\nu}^2)}
 {}_2\text{F}_2\!\left[\begin{matrix}1,1\\\fr32-\ii \widetilde{\nu},\fr32+\ii \widetilde{\nu}\end{matrix}\middle| 2\ii z\right], 
\end{align}
where $\text{J}_{\rho}$ is Bessel function of the first kind, ${}_2\text{F}_2$ a generalized hypergeometric function, and the coefficients $\be_\aa$ defined by
\begin{align}
 \be_\aa(\wt\nu)\equiv\frac{\sqrt{2\pi^3}\cos(\fr{\ii\aa\pi\widetilde{\nu}}{2}-\fr\pi4)}{4\cos(\ii\aa\pi \widetilde{\nu})\sin(\ii\aa\pi \widetilde{\nu})} . 
\end{align}

Cosmological collider signals are cleanest in the squeezed limit, which is closely related to the late-time limit $z\equiv-k\tau\to 0$ \cite{Arkani-Hamed:2015bza,Chen:2015lza,Assassi:2012zq}. It is thus useful to examine the late-time expansion of the mixed propagator $\Di(z)$, which has the following form:
\begin{equation} 
  \Di(z)=\sum_{\sA=0,\pm}\sum_{n=0}^\infty d_n^\sA z^{\Delta^\sA+n}.
\end{equation}
Here we have introduced three scaling dimensions $\Delta^{\sA}$ with $\sA\in\{0,\pm 1\}$, whose values are $\Delta^0=2$ and $\Delta^\pm=\tfrac32\pm\ii\nv$. They describe the three independent leading fall-off behaviors of the mixed propagator in the late-time limit, among which $\Delta^\pm$ are sources of non-analytic CC signals in the squeezed limit and $\Delta^0$ provides the analytic background. Every fall-off is followed by a series of overtones described by $n$. 

The Taylor coefficients $d_n^\sA$ can be found explicitly as:
\begin{align}
 d_{2j}^{\pm}&=
 \frac{(-1)^j\be_\pm(\wt\nu)}{2^{\pm\ii\nv+2j}j!\Gamma(1+j\pm\ii\nv)},
  ~~~d_{2j+1}^{\pm} =0,\n\\
 d_n^0&=
 \sum_{j=0}^{\lfloor n/2\rfloor}
 \frac{-(-\ii)^n/2}{(n-2j)!}\displaystyle\prod_{\ell=0}^{j}
 \FR{1}{(n-2\ell+\tfrac12)^2+\nv^2}.
 \label{eq:analyticcoeff}
\end{align}
The mixed propagator also satisfies a simple differential equation, which is a direct consequence of the mode equation:
\begin{equation}\label{eq:ode}
 \big(z^2\partial_z^2-2z\partial_z+z^2+\fr94 +\nv^2\big)\Di(z)
 =-\tfrac12z^2e^{-\ii z}.
\end{equation}
The equation is useful as it leads to recursion relations for $d_n^\sA$. More importantly, it gives rise to recursions for the mixed propagator itself in the Mellin space, as shown below.

\begin{figure}
\includegraphics[width=0.3\textwidth]{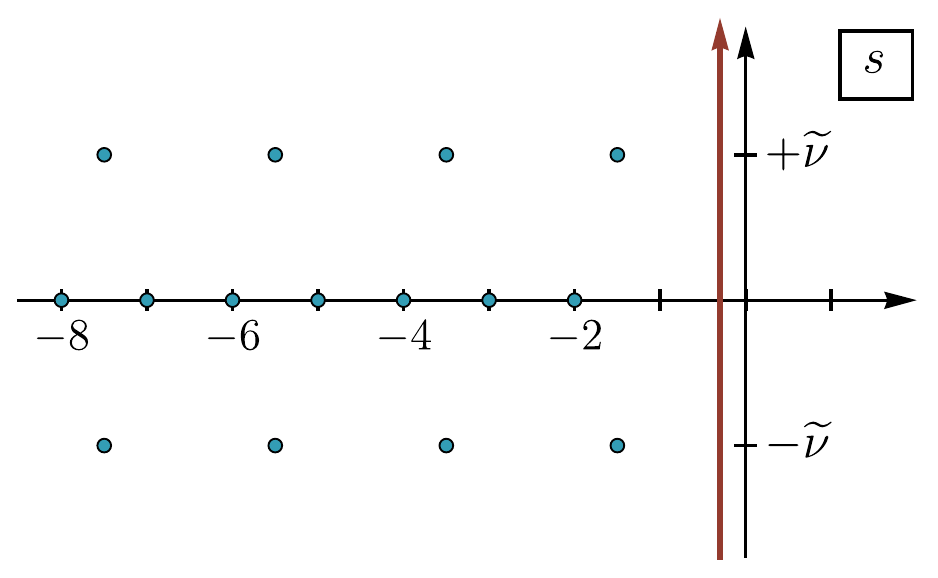}
\caption{The pole structure of Mellin-space mixed propagator $\M(s)$. The two series of poles on the upper and lower half plane correspond to $s=-\Delta^{\pm}-2n=-3/2-2n\pm\ii\wt\nu$ $(n\in\mathbb{N})$ which leads to CC signals in the squeezed limit, while the pole on the real axis, $s=-\Delta^0-n=-2-n$, provides the analytic background. The red line shows the Mellin integral contour.}
\label{fig_Mellin_poles}
\end{figure}

\noindent\textbf{\textit{Mellin-Barnes representation and the result.---}}
The bispectrum (\ref{eq_BispectrumTimeIntegral}) is a time integral over a product of three mixed propagators. To finish it, we take the inverse Mellin transform of the mixed propagator $\Di(z)$:
\begin{align}\label{eq:Mdef}  
 \Di(z)=&\int_{c-\ii\infty}^{c+\ii\infty}\FR{\di s}{2\pi\ii}\M(s)z^{-s}.
\end{align}
Here, $\M(s)$ is the mixed propagator in the Mellin space and can be worked out explicitly as
\begin{align}\label{eq:M} 
 &\M(s)=e^{-\ii\pi s/2}\Big[2^{s-2}\sqrt\pi\,
 \Gamma(\fr{s}2+\fr34+\fr{\ii\nv}{2})
 \Gamma(\fr{s}2+\fr34-\fr{\ii\nv}{2})\n \\ 
 & \times\big(\ii+\sech(\pi\nv)+\cot(\pi s)\big)-\fr14 \Gamma(s+1) \Xi_{\wt\nu}(s)\Big],
\end{align}
where the function $\Xi_{\wt\nu}(s)$ is defined via a single-layer sum:
\begin{align}\label{eq:Xi}
 \Xi_{\wt\nu}(s)=&\sum_{n=0}^\infty
 \frac{(\tfrac12+\ii\nv)_n(\tfrac12-\ii\nv)_n}{2^n n! (-s)_n}
 \big[\psi(n+1)+\psi(n-s)\n\\
 &-\psi(n+\tfrac12+\ii\nv)-\psi(n+\tfrac12-\ii\nv)+\log2\big]. 
\end{align}
The Mellin-space mixed propagator $\M(s)$ is meromorphic and has left poles at $s=-\Delta^0-n$ and $s=-\Delta^\pm-2n$ with $n\in\mathbb{N}$ whose residues are simply $d_n^\sA$ in (\ref{eq:analyticcoeff}), as shown in Fig.\;\ref{fig_Mellin_poles}. 

Also useful is the Mellin transform of the differential equation (\ref{eq:ode}), which in the Mellin space gives rise to a recursion relation: 
\begin{equation}\label{eq:Mrec}
 \M(s+2)+\big[(s+\tfrac32)^2+\nv^2\big]\M(s)
 =\tfrac12\Gamma(s+2)e^{-\ii\pi s/2}.
\end{equation}
This second-order recursion makes it convenient to generate values of $\M(s)$ for actual computation of the bispectrum $\J$, on which we shall comment later.

With the Mellin-Barnes representation, the time integral in (\ref{eq_BispectrumTimeIntegral}) can be finished directly, leading to a conservation condition for the three Mellin variables, which in turn eliminates one layer of Mellin integral. As a result, the bispectrum $\J$ is turned into a two-layer Mellin integral:
\begin{equation}\label{eq:MB}
 \J= \int \FR{\di s_1\di s_2}{(2\pi\ii)^2}\,
 \M(s_1)\M(s_2)\M(-3-s_1-s_2)r_1^{-s_1}r_2^{-s_2}.
\end{equation}

Since we work in the region $0<r_1\leq r_2< 1$, we can finish the two Mellin integrals by closing both contours from the left side and thereby picking up all left poles at 
$s_{1,2}=-\Delta^{\sA_{1,2}}-n_{1,2}$ and collect residues $d_{n_i}^{\sA_i}$ $(i=1,2)$. The result is:
\begin{align} 
  \mathcal{J}= 
 \sum_{\mathsf A_2, n_2}\bigg[\sum_{\mathsf A_1, n_1}\M(\Delta^{\mathsf A_1}+\Delta^{\mathsf A_2}+n_1+n_2-3) \prod_{i=1}^2 d_{n_i}^{\mathsf A_i} r_i^{\Delta^{\mathsf A_i}+n_i}\bigg], 
\label{eq_BispectrumDoubleSeries} 
\end{align}
where the summation range is $\sA_1,\sA_2\in\{0,\pm1\}$ and $n_1,n_2\in\mathbb{N}$.  
This is the final result for the weakly mixing bispectrum, which is a double series. A few comments are in order:

First, for our choices of physical region $0<r_1\leq r_2\leq 1$, the above summations should be performed in a specific order: first the sums over $n_1$ and $\sA_1$, and then $n_2$ and $\sA_2$. The $n_1$-sum is convergent when $|r_1|<1$. After finishing the $n_1$-sum, we need to sum over three fall-off branches labeled by $\sA_1=0,\pm1$. Then, we can proceed to the next layer summation over $n_2$, which has a radius of convergence $|r_2|<1+r_1$ since the nearest singularity to the origin is at $r_2=-1-r_1$ which is the so-called total-energy singularity \cite{Arkani-Hamed:2018kmz}. Finally we finish $\sA_2$-sum. This summation order is important: if we do not sum over $\sA_1$ first but directly proceed to $n_2$-sum, we will get spurious divergences at $r_2=1-r_1$ which is the well-known (spurious) folded singularity.

Second, the coefficients of the above series involve only the residues $d_{n_i}^{\sA_i}$ given in (\ref{eq:analyticcoeff}) and the Mellin-space mixed propagator $\M(s)$ in (\ref{eq:M}). All these objects are written in terms of finite sums and products of simple functions, with the only exception of the function $\Xi_{\wt\nu}(s)$ in $\M(s)$, which is generally a fast convergent series. Therefore, the series (\ref{eq_BispectrumDoubleSeries}) can be readily evaluated. One caveat is that, when $s$ takes a nonnegative integer value $N$, the expression (\ref{eq:M}) for $\M(s)$ contains individually divergent terms which cancel out and leave a finite $\M(N)$. This is not a problem in principle but makes (\ref{eq:M}) inconvenient for practical computations. In the End Matter, we provide alternative expressions $\M(N)$ for $N\in\mathbb{N}$ that are finite and easy to evaluate.  

Furthermore, the recursion relation (\ref{eq:Mrec}) allows us to reduce the evaluation of $\M(s)$ on infinitely many points shown in (\ref{eq_BispectrumDoubleSeries}) to 8 independent points:
\bge
  s\in\{0,1,\pm2\ii\wt\nu,\fr12\pm\ii\wt\nu,\fr32\pm\ii\wt\nu\}.
\ede
Practically, one can use (\ref{eq:M}) and corresponding formulae in End Matter to evaluate $\M(s)$ on these points, and then use the recursion to generate other $\M(s)$ to desired order.

With the above prescription on the summation order, the series (\ref{eq_BispectrumDoubleSeries}) is convergent for all points in the interior of physical region with $0<r_1\leq r_2\leq 1$ and $r_1+r_2>1$, except for the equilateral point $r_1=r_2=1$. Meanwhile, the convergence speed drops for kinematics close to the squeezed limit $r_1\ll 1$ and the equilateral limit. Fortunately, the Mellin representation is flexible enough to generate similar series with improved convergence in these regions, which we describe in the following.

\noindent\textbf{\textit{Squeezed and equilateral limits.---}}
The squeezed limit is most interesting for CC physics as the clock or power-law signal is most prominent in this region. With our choice of kinematics, the squeezed limit has $0\leq 1-r_2\leq r_1\ll 1$. This hierarchy suggests us to develop a series in $r_1$ and $1-r_2$ to improve the convergence in this region. This series is given in the End Matter and here we present the result for isosceles triangles with $r_2=1$.

Setting $r_2=1$ in (\ref{eq_BispectrumTimeIntegral}) and expanding in small $r_1$ inspire us to introduce the Mellin-Barnes representation of the squared mixed propagator as:
\bge
  \H(s)\equiv\int_0^\infty\di z\,z^{s-1}\Di(z)^2,
\label{eq_Hs}
\ede
$\mathcal{H}(s)$ can be expressed analytically as a convergent series whose explicit form is given in End Matter. Then, in terms of $\mathcal{H}(s)$, it is straightforward to obtain a simple expansion of $\J$ in small $r_1$:
\begin{align}\label{eq:iso}
 \J(r_1,1)=\sum_{n,\sA} d_n^\sA 
 \H(n+\Delta^\sA-3)r_1^{\Delta^\sA+n} .
\end{align}
In particular, the overall amplitude of the leading CC signal in $r_1\ll 1$ comes from $n=0$ and $\sA=\pm$. In the large-mass limit $\wt\nu\gg 1$, this part reproduces known result \cite{Xianyu:2025lbk}:
\begin{equation}\label{eq:shapeLarge}
 \mathcal{S}_\text{cc}\sim
 \frac{\sqrt\pi\,\rho^3\lambda_3}{32\sqrt{P_\zeta}}
 \nv^{-5/2}e^{-\pi\nv}r_1^{1/2}
 \cos\big(\nv\log\fr{r_1}4-\fr{\pi}4\big).
\end{equation}

It happens that the series (\ref{eq:iso}) actually converges at $r_1=1$ and thereby gives an analytical expression for the bispectrum at the equilateral point $r_1=r_2=1$:
\bge
\label{eq_equilateral}
  \J(1,1)=\sum_{n,\sA} d_n^\sA \H(n+\Delta^\sA-3).
\ede
This is a function of mass $\wt\nu$ only. In the large-mass limit, it reduces to $\J(1,1)\sim\frac{-\ii}{108\nv^6}[1+\fr{21}{4\nv^2}+O(\nv^{-4})]$, whose leading $1/m^6$ behavior is expected from integrating out the $\si$ field \cite{DuasoPueyo:2025lmq,Li:2026xhw,Dorigoni:2026mbh}.

\noindent\textbf{\textit{Nonperturbative mixing.---}} As pointed out in \cite{Chen:2009zp}, a weak mixing between $\phi$ and $\si$ is required for its perturbative treatment but not a model-building requirement. Ref.~\cite{An:2017hlx,Iyer:2017qzw} find that large mixing leads to new bispectrum shapes and Ref.~\cite{Huenupi:2026abj,Huenupi:2026aqc,Pinol:2026xnl,Wang:2026lff,Belrhali:2026uxn,Pinol:2026hey} find the propagator with general mixing. Remarkably, the strategy of Mellin-space mixed propagator of this work generalizes readily to the nonperturbative mixing case. For finite $\rho$, we directly work with the mixing $\la\varphi_{\bm k}(0)\si_{-\bm k}(\tau)\ra'\equiv \Di_\rho(-k\tau)/k^3$, where $\Di_\rho(z)$ is the nonperturbative mixed propagator which reduces to $\Di_\rho(z)=\rho\Di(z)+\order{\rho^3}$ in the weak mixing limit. The nonperturbative mixed propagator $\Di_\rho(z)$ also allows for an analytical Mellin-Barnes represenation $\M_\rho(s)$, given in the End Matter. The function $\M_\rho(s)$ also has left poles at $s=-\Delta_\rho^0-n$ and $s=-\Delta_\rho^{\pm}-2n$, corresponding to analytic and CC parts in the late-time limit, respectively. Importantly, while $\Delta_\rho^0=2$ is as before, the clock dimensions are shifted to $\Delta_\rho^\pm=\fr32\pm\ii\wt\nu_\rho$ with $\wt\nu_\rho=\sqrt{m^2+\rho^2-9/4}$. Denoting the residue of $\M_\rho(s)$ at $s=-\Delta_\rho^\sA-n$ as $d_{\rho,n}^\sA$, the above treatment applies directly in this case, leading to the dimensionless bispectrum for the nonperturbative mixing case:
\begin{align}\label{eq:strongseries}
 \J_\rho={}&\sum_{\sA_2,n_2}\bigg[
 \sum_{\sA_1,n_1}
 \M_\rho (\Delta_{\rho}^{\sA_1}+\Delta_{\rho}^{\sA_2}+n_{1}+n_2-3) \n\\
 &\times
 \prod_{i=1}^2
 d_{\rho,n_i}^{\sA_i}r_i^{\Delta_\rho^{\sA_i}+n_i}\bigg].
\end{align}
This is the final result for the bispectrum with general mixing strength.
Like before, the sum should be finished in a particular order to ensure convergence in the interior of physical region except for the equilateral point, where we can use (\ref{eq_equilateral}) with the replacement $d_n^\sA,\Delta^\sA,\H(s)\to d_{\rho, n}^\sA,\Delta_\rho^\sA,\H_\rho(s)\equiv\int_0^\infty \di z\,z^{s-1}\Di_\rho^2(z)$.

It remains to mention that, in the nonperturbative mixing case, the conversion from $\J_\rho$ to the shape function $\mathcal{S}_\rho$ requires a correction due to the finite shift of the power spectrum:
\begin{equation}\label{eq:strongshape}
 \mathcal S_\rho=-\frac{\lambda_3 }
 {\pi P_{\zeta,\rho}^{1/2}(2C_\rho)^{3/2}}\,
 \frac{\Ima\J_\rho}{r_1r_2},
\end{equation}
where the power spectrum $P_{\zeta,\rho}=C_\rho/(2\pi^2\dot\phi_0^2)$ and $C_\rho \equiv k^3\langle\varphi_{\bm k}(0)\varphi_{-\bm k}(0)\rangle'$ whose explicit form is given in the End Matter.

\noindent\textbf{\textit{Discussion.---}}
In this Letter, we combine two existing techniques, the mixed propagator and PMB representation, to provide convergent series solutions to triple-massive-exchange bispectrum, resolving a longstanding challenge in the field. We have tested all series solutions against direct numerical integration and found nice agreement. 

Our full solution gives an exact analytic answer for this model at general mass, mixing strength, and momentum configuration. It recovers known results in different limiting cases and provides analytic interpolations between them. These include the original weakly mixed QSFI models in the lighter \cite{Chen:2009we,Chen:2009zp,Baumann:2011nk} and heavier mass regimes \cite{Noumi:2012vr,Arkani-Hamed:2015bza}; the single-field limit with an effective sound speed in the heavy-spectator and weak-coupling regime \cite{Tolley:2009fg,Achucarro:2010jv,Gong:2013sma}; the strongly mixed effective single-field regime \cite{Baumann:2011su}; and the strong effective mass/mixing QSFI regime \cite{Belrhali:2026uxn}.
The solution interpolates analytically between complementary- and principal- series behavior across the critical effective mass $3H/2$, with a logarithmically modified squeezed-limit shape at the transition \cite{Chen:2009zp}, between weak and strong mixing across the boundary of the perturbative convergence region $|\rho|<m$, and between squeezed and equilateral configurations. Note that, even in some of these limits, previous results were available only for simpler coupling terms, or only in terms of numerical amplitudes or shapes, or only at leading order. A detailed exploration of the physics across all phases of this and other related models is now an interesting and tractable direction for future work.

Regarding numerical evaluation and data analyses, given the highly developed numerical strategies, the series may not always be the best way for numerical computation. Nevertheless, we do observe significant increase of evaluation speed with our series solutions with appropriate preprocessing of certain initial parameters. 
In addition, a full control over analytic properties of non-Gaussianities could provide valuable guidance on template design.

Regarding the analytic computational techniques, the method of Mellin-space mixed propagators can be broadly applied.
Another important observation is that, for practical use, a series solution needs not to be absolutely convergent. Rather, a suitably chosen summation order can significantly improve the convergence. Indeed, without the order prescription, (\ref{eq_BispectrumDoubleSeries}) converges only in the unphysical region with $r_1+r_2<1$, but the summation order prescription enlarges the convergence domain to almost entire physical region. This strategy can be used to improve many previously obtained series solutions to cosmological correlators such as the (massive) family-tree representation \cite{Xianyu:2023ytd,Fan:2024iek,He:2024olr,Fan:2025scu,Liu:2024str,Xianyu:2025lbk} and we believe this is an interesting direction to pursue.
Furthermore, even though our method is a direct computation, its result can be used as a resource for bootstrap studies of cosmological correlators \cite{Liu:2024xyi,Baumann:2026atn,Arundine:2026myr}. In particular, our result of weak mixing bispectrum takes a very compact form, with two independent momentum ratios and three layers of summation, supporting the transcendental-weight conjecture in \cite{Xianyu:2025lbk}.  

\begin{acknowledgments}
Claude Fable and GPT-6 Astra were used to assist with the analysis and the authors are responsible for all results.
ZX is supported by NSFC under Grants No.\ 12275146, No.\ 12247103, No.\ 12521007, and the Dushi Program of Tsinghua University. 
\end{acknowledgments}

\onecolumngrid
\begin{center}
    \textbf{End Matter}
\end{center}
\twocolumngrid
\noindent\textbf{\textit{Mellin propagator at integer values.---}}
In the series expression for the bispectrum (\ref{eq_BispectrumDoubleSeries}), the coefficients involve a Mellin-space mixed propagator $\M(s)$ whose argument may take all nonnegative integer values $N$, at which points the original expression (\ref{eq:M}) develops spurious poles for individual terms but the divergences cancel out. After canceling these divergences, we get finite values for $\M(N)$. Here we spell out these finite values directly:
\begin{align} 
 &\M(N)=(-\ii)^N\Big\{2^{N-2}\sqrt\pi\Gamma(q_+)\Gamma(q_-)
 \Big[\ii+\sech(\pi\nv)\n\\ 
 &+\frac{\log2-\psi(N+1)+\tfrac12[\psi(q_+)+\psi(q_-)]}{\pi}
 \Big]-X_N\Big\},
 \label{eq:MN}
\end{align} 
where $q_\pm\equiv\fr{N}2+\fr34\pm\fr{\ii\nv}{2}$ and $X_N$ is defined by:
\begin{align}\label{eq:XN} 
 X_N=&\sum_{n=0}^N
 \frac{(-1)^n(N-n)!(\tfrac12+\ii\nv)_n(\tfrac12-\ii\nv)_n}
 {\,n!\,2^{n+2}}\n\\
 &\times[A_n+2\psi(N-n+1)-\psi(N+1)]\n\\
 &+\sum_{n=N+1}^\infty
 \frac{(-1)^N(\tfrac12+\ii\nv)_n(\tfrac12-\ii\nv)_n}
 {2^{n+2} n!(n-N-1)!}\big[\psi^{(1)}(n-N)\n\\
 & +(H_N-H_{n-N-1})\big(A_n+\psi(n-N)\big)\big].
\end{align}
Here we further used the trigamma function $\psi^{(1)}$ and the harmonic number $H_n=\sum_{j=1}^n 1/j$ for $n\in\mathbb{Z}_+$ and $H_0=0$. Furthermore, the function $A_n$ is defined by
\bge
A_n\equiv\psi(n+1)-\psi(n+\tfrac12+\ii\nv)
                 -\psi(n+\tfrac12-\ii\nv)+\log2.
\ede 
For practical numerical implementation, it may be more convenient to compute only $\M(0)$ and $\M(1)$ from these expressions, and then use the recursion (\ref{eq:Mrec}) to generate $\M(N)$ for $N\geq2$.

\noindent\textbf{\textit{Improved series in the squeezed limit.---}}  
The squeezed-limit series in (\ref{eq:iso}) include the squared mixed propagator in the Mellin-space $\H(s)$ defined in (\ref{eq_Hs}). It is possible to derive a convergent series for $\H(s)$ by considering a slightly more general object $F_s(r)=\int_0^\infty\di z\,z^{s-1}\Di(rz)\Di(z)$: We can expand $F_s(r)$ around $r=2/3$, a convenient but slightly arbitrary choice, whose $\ell$-th Taylor coefficient can be easily worked out to be $\T_\ell(s)$ given below. Then, we can use this expansion to express $\H(s)=F_s(1)$ as:
\begin{align} \label{eq:Hseries}
 \H(s)&=\sum_{\ell=0}^\infty3^{-\ell}\T_{\ell}(s),\\
 \T_{\ell}(s)&=\sum_{\sA=0,\pm}\sum_{m=0}^\infty
 d_m^{\sA}\M(s+\Delta^{\sA}+m)\n\\
 &\times \binom{\Delta^{\sA}+m}{\ell}
 (\fr23)^{\Delta^{\sA}+m-\ell},
\end{align}
where $\binom a b$ is the binomial factor.  
Similarly, we can derive a complete series expansion for $\J$ in the squeezed limit $0\leq 1-r_2\leq r_1\ll 1$ by expanding the integrand of (\ref{eq_BispectrumTimeIntegral}) with small $r_1$ and small $1-r_2$. The resulting series is:
\begin{align}\label{eq:softgeneral}
 \J(r_1,r_2)=&\sum_{\sA,n}d_n^{\sA}r_1^{\Delta^{\sA}+n}
 \sum_{k=0}^\infty(r_2-1)^k\n\\
 &\times\sum_{\ell=k}^\infty
 \binom\ell k 3^{k-\ell}\T_{\ell}(\Delta^{\sA}+n-3).
\end{align}
In particular, when $r_2=1$, this series reduces to the isosceles series given in (\ref{eq:iso}).

\noindent\textbf{\textit{Nonperturbative mixing.---}} Here we list explicit expressions for all variables appeared in the series solution for nonperturbatively mixed bispectrum in (\ref{eq:strongseries}) and (\ref{eq:strongshape}). First, the power spectrum in this case is given by \cite{Huenupi:2026abj}: 
\begin{align}\label{eq:strongpower}
 C_\rho =\frac12\left|
 \frac{\Gamma(\tfrac34+\tfrac{\ii\nv_\rho}{2})
       \Gamma(\tfrac34-\tfrac{\ii\nv_\rho}{2})}
 {\Gamma(\tfrac34+\tfrac{\ii(\rho+\nv_\rho)}2)
  \Gamma(\tfrac34+\tfrac{\ii(\rho-\nv_\rho)}2)}
 \right|^2.
\end{align}
Then, the Mellin-space nonperturbatively mixed propagator $\M_\rho(s)$ has the following series representation: 
\begin{align}\label{eq:strongM}
 &\M_\rho(s)=\FR{e^{-\ii\pi s/2}}{2^s}\Gamma(s)
 \sum_{\aa=\pm1}w_\aa\sum_{h=0}^4 W_\aa^h \sum_{n=0}^\infty
 \frac{(\alpha^h)_n(c_\aa^h-\beta^h)_n}
 {(c_\aa^h)_n n!} \nonumber\\
 &
 \times B(p_\aa^h+n+1,\alpha^h+s-p_\aa^h-q_\aa^h-1)
 \nonumber\\
 &\times\sum_{\ell=0}^\infty
 \frac{(s)_\ell(\alpha^h+s-p_\aa^h-q_\aa^h-1)_\ell}
 {2^\ell\ell!(\alpha^h+s-q_\aa^h+n)_\ell }.
\end{align}
Here $B(u,v)=\Gamma(u)\Gamma(v)/\Gamma(u+v)$, and there are many variables involved, which we list in the following and explain their origin in a forthcoming companion work. First, the two coefficients $w_\aa$ are:
\begin{align}\label{eq:strongweight}
 w_\aa={}& 
 \frac{-\ii\aa e^{-\pi\aa\rho/2}\Gamma(\tfrac34-\tfrac{\ii\nv_\rho}{2})
       \Gamma(\tfrac34+\tfrac{\ii\nv_\rho}{2})
       \Gamma(\tfrac12-\tfrac{\ii\aa\rho}{2})}
 {8\sqrt\pi\Gamma(-1+\tfrac{\ii\aa\rho}{2})\Gamma(\tfrac34-\tfrac{\ii\aa\rho+\ii\nv_\rho}2)
  \Gamma(\tfrac34-\tfrac{\ii\aa\rho-\ii\nv_\rho}2)},
\end{align} 
and the five rows are
\begin{equation}\label{eq:strongrows}
\begin{array}{c|ccc}
 h&p_\aa^h&q_\aa^h&c_\aa^h\\\hline
 0&-2+\ii\aa\rho/2&-1-\ii\aa\rho/2&1+\ii\aa\rho\\
 1&-1+\ii\aa\rho/2&-1-\ii\aa\rho/2&1+\ii\aa\rho\\
 2&-1+\ii\aa\rho/2&-2-\ii\aa\rho/2&1+\ii\aa\rho\\
 3&\ii\aa\rho/2&-1-\ii\aa\rho/2&2+\ii\aa\rho\\
 4&-1+\ii\aa\rho/2&-1-\ii\aa\rho/2&2+\ii\aa\rho
\end{array}
\end{equation}
The remaining parameters and weights are
\begin{align}\label{eq:strongrowweights}
 (\alpha^h,\beta^h)&=
 \begin{cases}
 (\tfrac12-\ii\nv_\rho,\tfrac12+\ii\nv_\rho),&h=0,1,2,\\
 (\tfrac32-\ii\nv_\rho,\tfrac32+\ii\nv_\rho),&h=3,4,
 \end{cases}\nonumber\\
 W_\aa^0&=1,~~
 W_\aa^1=\frac{4\nv_\rho^2+1}{\aa\rho(\aa\rho+2\ii)}, ~~
 W_\aa^2 =-\frac{\aa\rho-2\ii}{\aa\rho+2\ii},\nonumber\\
 W_\aa^3&=2W_\aa^4=-\frac{8(1/4+\nv_\rho^2)}
 {(1+\ii\aa\rho)\aa\rho(\aa\rho+2\ii)}.
\end{align}
The sum in (\ref{eq:strongM}) is to be performed in the indicated order to ensure convergence. Also, unlike perturbative case, $\M_\rho(s)$ is already nonsingular for all $s\in\mathbb{Z}_+$. The only point that needs extra care is $s=0$, where we can use the following formula:
\begin{equation}\label{eq:strongMzero}
 \M_\rho(0)=\frac{\rho C_\rho-\M_\rho(2)}{m^2}.
\end{equation}
Finally, we need the Taylor coefficients $d_{\rho,n}^\sA$ of late-time expansion of $\Di_\rho(z)$, or, equivalently, the residues of left poles of $\M_\rho(s)$. They can be generated from the following leading coefficients:
\begin{align}\label{eq:strongseeds}
 d_{\rho,0}^0&=-\frac{\rho C_\rho}{m^2+\rho^2-2},\nonumber\\
 d_{\rho,1}^0&=\frac{\rho}{2(m^2+\rho^2)}
 \left[\ii+\frac{\cosh(\pi\rho)-1}{\cosh(\pi\nv_\rho)}\right],
 \nonumber\\
 d_{\rho,0}^{\bb}&=-4(2\ii)^{\Delta_\rho^{\bb}}
 \Gamma(-2\ii\bb\nv_\rho)\nonumber\\
 &\quad\times\sum_{\aa=\pm1}
 \frac{w_\aa\Gamma(1+\ii\aa\rho)}
 {\aa\rho(\aa\rho+2\ii)
  \Gamma(\tfrac12+\ii\aa\rho-\ii\bb\nv_\rho)}.
\end{align}
Coefficients $d_{\rho,n}^\sA$ with $n>0$ are generated with the following matrix:
\begin{equation}\label{eq:strongtransfer}
 T(q)=\frac{
 \begin{pmatrix}
 -q(q-3)-m^2&-\rho(q-3)\\
 \rho q&-q(q-3)
 \end{pmatrix}}
 {q(q-3)[(q-\tfrac32)^2+\nv_\rho^2]}.
\end{equation}
Then, let $e_2^{\mathsf T}=(0\ \ 1)$ and we have:
\begin{align}\label{eq:strongres}
 {}&d_{\rho,2j}^0=C_\rho e_2^{\mathsf T}
 T(2+2j)T(2j)\cdots T(2)\binom10,
 \nonumber\\
 {}&d_{\rho,2j+1}^0=d_{\rho,1}^0 e_2^{\mathsf T}
 T(3+2j)\cdots T(5)\binom{-m^2/(3\rho)}1,
 \nonumber\\
 {}&d_{\rho,2j}^{\bb}=d_{\rho,0}^{\bb}e_2^{\mathsf T}
 T(\Delta_\rho^{\bb}+2j)\cdots T(\Delta_\rho^{\bb}+2)
 \binom{\rho/\Delta_\rho^{\bb}}1,
 \nonumber\\
 {}&d_{\rho,2j+1}^{\bb}=0,
\end{align}
with $j\in\mathbb{N}$ and that empty product is defined to be 1.

\bibliographystyle{utphys}
\bibliography{CosmoCollider}

\end{document}